# Ultrasound Detection Arrays

# Via Coded Hadamard Apertures

Evgeny Hahamovich and Amir Rosenthal

*Abstract*— In the medical fields, ultrasound detection is often performed with piezoelectric arrays that enable one to simultaneously map the acoustic fields at several positions. In this work, we develop a novel method for transforming a single-element ultrasound detector into an effective detection array by spatially filtering the incoming acoustic fields using a binary acoustic mask coded with cyclic Hadamard patterns. By scanning the mask in front of the detector, we obtain a multiplexed measurement dataset from which a map of the acoustic field is analytically constructed. We experimentally demonstrate our method by transforming a single-element ultrasound detector into 1D arrays with up to 59 elements.

*Index Terms*—Ultrasonic imaging, Ultrasonic transducers

## I. INTRODUCTION

Medical ultrasound imaging, or ultrasonography, conventionally relies on piezoelectric transducers to generate and detect ultrasound [1]. While early implementations were based on a focused single-element transducer that was mechanically scanned to produce 2D images, the introduction of transducer arrays has enabled scan-free formation of 2D and 3D images via electric beamforming [2],[3]. Since the measurement dataset is limited by the number of transducer elements, it is generally desired to maximize the number of elements to optimize image contrast and resolution.

In addition to conventional ultrasonography, piezoelectric array transducers have played a major role in the development of optoacoustic tomography (OAT). In OAT, ultrasound is generated within the imaged specimen via laser-pulse irradiation and subsequent thermal expansion, where the transducers are used only to detect the acoustic signals [4]. Image formation in OAT is achieved via tomographic inversion algorithms that are performed on the digitally voltage signals from the detectors [5].

The emergence of OAT as a tool for biomedical research and clinical diagnosis leads to new requirements in ultrasound detection that may not always be addressed by piezoelectric technology. In OAT, a wide tomographic view is needed to avoid loss of lateral resolution [6], whereas conventional linear piezoelectric arrays have acceptance angles of merely ±20 degrees [7]. Also, reducing the array pitch below 100 μm involves technological challenges that have not been fully resolved. As a result, array-based 3D optoacoustic is generally not performed at resolutions significantly better than 100 μm and curved detector arrays are often used, leading to coupling challenges in clinical applications, in which the detectors need to be in close contact with the skin. In addition, the signals in OAT are considerably weaker than in ultrasonography, necessitating the use of large-area piezoelectric elements to maximize sensitivity, which limits the number of elements in the array. Advanced OAT systems today employ up to 512 elements and can produce 3D images from a single laser excitation pulse, albeit with low visual quality [8]. Improvement in contrast and reduced image artifacts are often achieved by mechanical scanning of the detector array.

Additional imaging modality in which ultrasound transducers are used in reception mode, rather than as transceivers, is ultrasound computer tomography (UCT) [9]–[12]. In UCT, two transducers are positioned on both sides of the imaged object, where one transducer generates the acoustic waves and the other transducer receives the waves that propagated through the object. UCT enables quantitative measurement of both the attenuation coefficient and the speed of sound. The method is mostly useful for soft tissue imaging and may require scan of the transducers to achieve proper imaging resolution [13].

In recent years, coded detection scheme was applied to increase the measurement dataset obtained by piezoelectric transducers without mechanically scanning the transducer. In [14] 3D ultrasonography was demonstrated with a single-element transducer by placing a random, rotating phase mask on the transducers surface. The mask records a dataset of acoustic measurements sufficiently large to reconstruct objects in 3D. In [15] reduction in the number of required detectors for 2D OAT reconstructions was achieved by incorporating random acoustic scatterers between the imaged object and the detector array. In both works, the random acoustic response of the added element, either mask or scatterers, was measured in advanced, and compressed-sensing algorithms were used for image reconstruction.

In addition to the random coding schemes of [14]-[15], structured multiplexing of the transmitted and detected acoustic fields have been demonstrated with transducer arrays. In transmission, Hadamard coding [16] and S-sequence [17], [18] coding were demonstrated by changing the transmitting elements in each ultrasound burst based on the coding matrix. In reception, multiplexed detection was demonstrated with capacitive micromachined ultrasonic transducers (CMUTs) by leveraging the bias of top-orthogonal-to-bottom-electrode of the CMUT architecture [19]. The advantage of multiplexed transmission and reception is the increased sensitivity it offers compared to unmultiplied schemes. However, the schemes of [16]-[19] neither increased the angular acceptance of the transducer nor reduced the number of detection elements needed to form an image.

In this work, we developed a novel method for spatial

### TABLE I
#### SUMMARY TABLE FOR WEIGHTS CONFIGURATIONS

| System Configuration | Possible Weight Values | SNR Improvement | Optimal weighing matrix |
|---|---|---|---|
| General case | $[-1{:}1]$ | $\sqrt{\dfrac{N}{\mathrm{tr}\left[\left(\mathbf{W}^{\mathrm{T}}\mathbf{W}\right)^{-1}\right]}}$ | General weighing matrix |
| Chemical balance (2-pan) | -1, 1 | $\sqrt{N}$ | Hadamard |
| Spring balance (1-pan) | 0, 1 | $\sqrt{\dfrac{(N+1)^2}{4N}}\approx\dfrac{\sqrt{N}}{2}$ | S-matrix |

*Summary of the configurations and the compatible weighing and SNR gain values. The presented SNR gain is in comparison to the SNR of a single detector measurement.*

multiplexing of ultrasound signals that is based on scanning coded-Hadamard-aperture (CHA) masks in front of large-area detectors. Each aperture in the CHA performs as a point-like detector with semi-isotropic sensitivity [20], thus effectively transforming the single detector into an array. In contrast to [14]-[15], our multiplexing approach is based on optimal codes that maximize sensitivity and may be analytically inverted without regularization, rather than on random arrangement of acoustic elements that required regularization to be inverted. Specifically, we use cyclic S-matrix coding in which all base function are obtained from cyclic shifts of the same code [21].

We experimentally demonstrate the capability of CHAs to transform a single-element ultrasound detector into a 1D array of up to 59 emulated detectors with a central frequency of 1 MHz by spatially filtering the incoming ultrasonic signal. The achieved sensitivity gain with respect to a single aperture fits the theoretical prediction of $\sqrt{N}/2$, where $N$ is the number of emulated detectors. In addition, our results indicate that the semi-isotropic sensitivity we previously obtained for a single aperture [20] is preserved for CHAs.

### II. THEORETICAL BACKGROUND

In this section we present the theoretical underpinning of spatial multiplexing.

#### A. Spatial multiplexing

Assuming the configuration of $N$ measurements via a single detector in an additive independent noise regime, where the additive noise $\mathbf{n}$ is composed of $N$ independent and identically distributed random variables with a normal distribution: $\mathbf{n} \sim N\left(\mathbf{0}, \sigma^2\mathbf{I}\right)$. One can perform a *spatially multiplexed* measurement of the $N$ elements of a vector $\mathbf{x}$ via a linear projection of the signals through a multiplicative weighing matrix $\mathbf{W}$, where $\mathbf{W}$ is an $N{\times}N$ weight matrix, which determines for each one of the $N$ measurements the sampled elements of the vector $\mathbf{x}$. Mathematically, described by

$$\mathbf{y} = \mathbf{Wx} + \mathbf{n}, \tag{1}$$

where $\mathbf{y}$ is the measurements vector. Assuming $\mathbf{W}$ is invertible,

the acoustic signals are recovered from the measurement via $\hat{\mathbf{x}} = \mathbf{W}^{-1}\mathbf{y}$, resulting with

$$\hat{\mathbf{x}} = \mathbf{x} + \mathbf{W}^{-1}\mathbf{n}. \tag{2}$$

The noise covariance matrix for this *spatially multiplexed* measurement is

$$\mathbf{K} = \sigma^2 \cdot \left(\mathbf{W}^{\mathrm{T}}\mathbf{W}\right)^{-1} \tag{3}$$

where T denotes the transpose operation. We denote the average mean square error (MSE) by

$$M = \frac{1}{N}\sum_{j=1}^{N} E\left[\left(\hat{x}_j - x_j\right)^2\right], \tag{4}$$

where $x_j$ is the $j^{th}$ element of the vector $\mathbf{x}$. For the described diagonal noise matrix, the average MSE is given by [21]

$$M = \sigma^2/N \cdot \mathrm{tr}\left[\left(\mathbf{W}^{\mathrm{T}}\mathbf{W}\right)^{-1}\right], \tag{5}$$

where tr denotes the trace operation.

A straightforward scenario of a *direct* measurement, $\mathbf{y} = \mathbf{x} + \mathbf{n}$, where a single element of the vector $\mathbf{x}$ is acquired per measurement can be represented via the identity matrix as a weighing matrix, $\mathbf{W} = \mathbf{I}$. The covariance matrix and the average MSE for the *direct* measurement are:

$$\mathbf{K}_{\mathrm{direct}} = \sigma^2\mathbf{I}, \tag{6}$$

$$M_{\mathrm{direct}} = \sigma^2. \tag{7}$$

Leading to an SNR gain of

$$G = \sqrt{N/\mathrm{tr}\left[\left(\mathbf{W}^{\mathrm{T}}\mathbf{W}\right)^{-1}\right]} \tag{8}$$

for the multiplexed measurement for a general case weighing matrix $\mathbf{W}$ in comparison to the direct measurement. The desired scenario of $G > 1$ is often referred to as *the multiplexing advantage* [22].

#### B. Optimal multiplexing codes

In imaging applications, the multiplexing matrix $\mathbf{W}$ is determined by the physics of the problem and its entries may generally accept any real value. However, when multiplexing is not a direct consequence of the underlying physics of the measurement system, it may be introduced in order to achieve the multiplexing advantage. In that case, $\mathbf{W}$ is not pre-determined by the physics, but rather designed. One could use a weighing matrix minimizing the measurement error. Common multiplexing matrix used in such a case is the Hadamard matrix, with the binary elements of $\pm 1$ [23]. For the Hadamard matrix $\mathbf{H}^{\mathrm{T}}\mathbf{H} = N\mathbf{I}$, leading to

$$\mathbf{K}_{\mathrm{H}} = \frac{\sigma^2}{N}\mathbf{I} \tag{9}$$

$$M_H = \sigma^2/N \tag{10}$$

$$G_{\mathrm{H}} = \sqrt{N} \tag{11}$$

This configuration in which the multiplexing matrix weight are

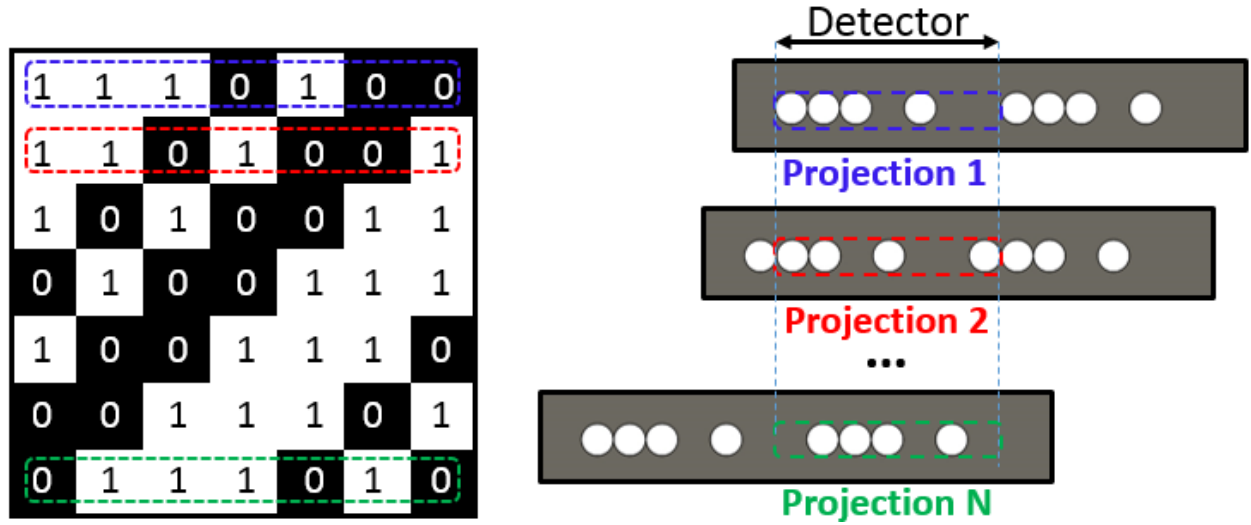

Fig. 1. Method illustration by a 7[th] order cyclic CHA. On the left, the 7x7 cyclic S-matrix and on the right the corresponding physical implementation with the CHA mask. The base functions are changed by shifting the mask while the shift distance per acquisition is the distance between the elements.

±1 is often referred to as a chemical-balance design [24]. When such an implementation is impractical, a simpler spring-balance design may be used, in which the two acceptable matrix entries are "0" and "1". Under this restriction, the optimal SNR gain is

$$G_s = \sqrt{(N+1)^2/(4N)} \qquad (12)$$

and may be achieved by a variant of the Hadamard matrix often referred to as the *S-matrix*. The properties of the chemical- and spring-balance approaches are summarized in Table I.

### C. Physical interpretation

In our specific work regime of ultrasound measurements, the noise term in (1) appears after the multiplexing operation since the multiplexing is not performed electronically or digitally on the detected signals, but rather acoustically before the signals reach the detector. Our model assumes that the dominant noise source is additive detector noise, which generally applies for ultrasound detection [25]. Other noise models, such as photon shot noise in optics should be analyzed differently [26].

Every acquired ultrasonic signal is a vector of time samples recorded per measurement, transforming the vector $\mathbf{x}$ into a matrix of size $N \times t$. Therefore, the formulation of (1) is changed to a matrix representation of $\mathbf{Y}$, $\mathbf{X}$ and $\mathbf{N}$:

$$\mathbf{Y}_{N \times t} = \mathbf{W}_{N \times N} \mathbf{X}_{N \times t} + \mathbf{N}_{N \times t} \qquad (13)$$

where $\mathbf{t}$ represents the time samples acquired per measurement and N represents the measurements base. As can be seen from the equation, each sample in time is multiplexed with the corresponding N measurement of the same time window. This can be also seen as a "slow" time which separate between the N consecutive measurements and a "fast" time which is the sampling time per acquisition.

Physical interpretation of the discussed weighing matrix for detection is: +1 elements of the weighing matrix correspond to signal detection, -1 elements correspond to invert phase detection of the signals and zeroes correspond to signals blocked by the mask. While choosing the invert phase option leads to the best possible SNR available by multiplexing, implementation of an inverse phase is technically more challenging, especially for wideband signals in which the accumulated phase over a given distance of each frequency is different.

### D. Cyclic codes

Generally, to perform spatial multiplexing for $N$ acoustic signals with an *S-matrix* one would need to produce a configurable mask with $N$ pixels whose transmission may be individually modified between "0" and "1" for each pixel. A single acoustic detector would then measure the signals for $N$ different configurations of the mask to produce the multiplexed signals. While in the field of optics, configurable spatial modulators are implemented via the use of digital micro-mirror devices [27], such a technology does not exist for ultrasound. Nonetheless, as demonstrated in [20], one may construct non-configurable acoustic masks that transmit the acoustic wave at specific location while blocking it in others. However, to produce a set of $N$ multiplexed signals with this approach, one would generally need to produce $N$ physical masks and switch between them, leading to impractical scanning durations when $N$ is large.

We propose to perform a spatially multiplexed acoustic measurement with a single mask by using a cyclic S-matrix in which each row vector is cyclically shifted by one element relative to the preceding row vector per measurement. Thus, the cyclic S-matrix is uniquely determined by the $N$ entries of any one of its rows. In this work, we use the quadratic residue construction algorithm [28], which allows one to construct S-matrices for every prime $N$ that may be presented in the form of $4m+3$, where $m \in \mathbb{N}$.

To physically implement a cyclic S-matrix, a single CHA

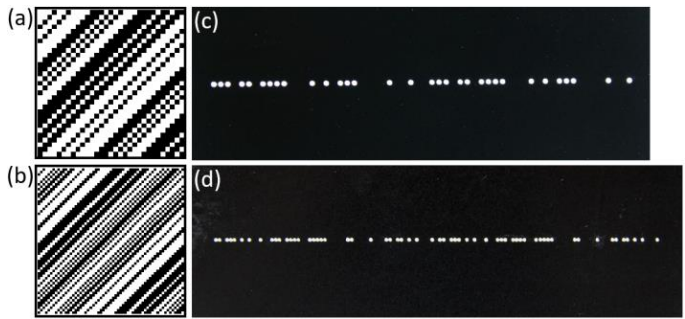

Fig. 2. The two CHA masks used for the experiment. Mask (c) is a photograph of the 31 elements mask with 1.5 mm pitch and 1 mm diameter and mask (d) is a photograph of the 59 elements mask with 1 mm pitch and 1 mm diameter. Respectably, the full base, cyclic, S-matrix generated by the masks is shown for N=31(a) and N=59 (b) with detected elements in white and blocked elements in black.

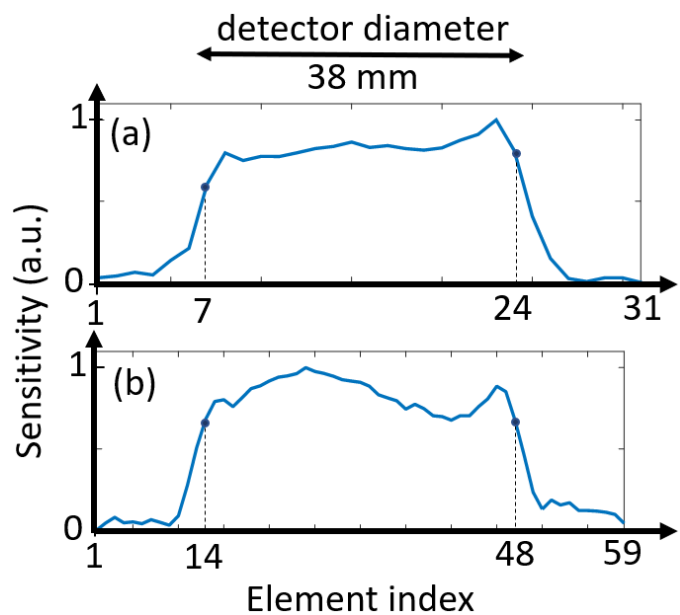

Fig. 4. Sensitivity of the virtual detectors for 1D virtual arrays with (a) 31 elements and (b) 59 elements. Active elements are 7-24 for (a) and 14-48 for (b). The active area shown in comparison to the detectors diameter.

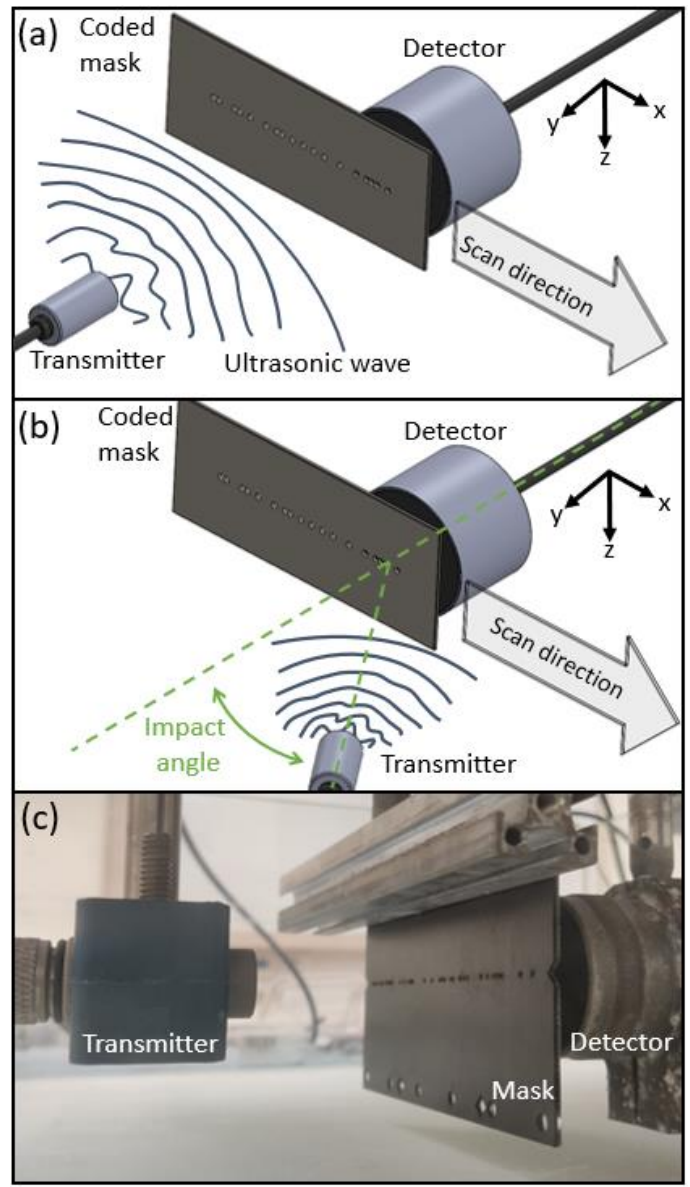

Fig. 3. Experimental setup. The field of an ultrasonic transmitter is recorder by a single detector, spatially filtered by a coded moving mask. Providing multiplexed recording of the transmitters field map by a virtual multi-element array. (a) illustrates the direct impact setup, (b) illustrates the angle impact setup and (c) is a photograph of the direct impact setup as illustrated in (a).

mask with 2N-1 elements is required, where the detector measures the transmission through only N of these elements at any given projection. To switch between neighboring rows of the S-matrix, the mask is shifted by a single cell with respect to the detector, which remains static during the measurement. Fig. 1 shows an illustration of the S-matrix and corresponding CHA mask for the case of N=7.

### III. METHODS

To evaluate the proposed approach, two cyclic CHA masks were manufactured by producing circular apertures in a thin acoustic absorber. Similar to [20], the acoustic absorber was composed of 0.53 mm layer of PORON 4701-30-25 foam bounded to a 1 mm thick polymethyl-methacrylate (PMMA) backing plate, whereas the apertures were produced via laser cutting. The first CHA mask coded an S-matrix with N=31 using circular apertures with a diameter of 1.5 mm, where the distance between the centers of the apertures was 2 mm. The second CHA mask coded an S-matrix with N=59, where both the aperture diameter and distance between the centers of neighboring apertures were 1 mm. Accordingly, the first mask corresponded to a virtual detector array detector with a length of 61.5 mm, whereas for the second mask the array length was 59 mm. Photographs of the manufactured masks and their corresponding S-matrices are shown in Fig. 2. Additionally, for comparison purposes, two single aperture masks were manufactured, both with the same element diameter as in the two CHA masks.

The measurement setup, shown in Fig. 3a and 3c, comprised of an ultrasound transmitter (Olympus, A303S) and receiver (Olympus, V392-SU) operating at 1 MHz with diameters of 12.7 mm and 38 mm, respectively. The transmitter was connected to an electric pulse generator (PicoPulser, Ultratek) and the receiver to a sampling card (M3i.4860-Exp., Spectrum Instrumentation) with 16-bit resolution. The CHA mask was positioned in front of the detector at a distance of approximately 1.5 millimeters, i.e. equal to the acoustic wavelength, to minimize the acoustic diffraction from the aperture to the detector. The mask was scanned in the *x* direction with *N* discrete steps equal to the distance between the mask elements. The mask step size in the x direction was 2 mm for the 31 elements mask and 1 mm for the 59 elements mask. The resulting N multiplexed measurements were used to calculate the acoustic signal of the N elements of the virtual detector array via $\hat{\mathbf{x}} = \mathbf{W}^{-1}\mathbf{y}$ demultiplexing. The measurements were performed inside a water tank with 100 Hz pulse repetition rate and averaged 16 samples per acquisition. The mechanical scan of the transducers and the mask were performed automatically by linear motorized stages (Zaber, T-LSR150B). One of the linear stages performed the linear scan of the mask in the *x* direction and two more stages enabled changing the relative position between the transmitter and the detector in *x*-*y* plane. Scan in the y direction was required to map the acoustic fileds for different depths and scan in the x derection was performed to align the transmitter and the detector during unifirmity measurement, described in section A.

Three sets of measurements were performed. First, sensitivity evaluation of the resulting detection array, second, acoustic field measurement and comparison to a single element detector in the time domain and via a 2D sensitivity map and third, angular response measurement of our technique.

### A. Uniformity and sensitivity

The first set of measurements was devoted to testing the uniformity and sensitivity of the virtual detector array. The transmitter and receiver were positioned coaxially 150 mm from each other, corresponding to a far-field measurement in which the signal amplitude over the detection area is approximately constant. A combined scan of the transmitter and the CHA mask was performed: The transmitter was placed in front of every one of the *N* virtual elements and for every transmitter position a multiplexed measurement was performed via a linear mask scan through *N* positions. The measurement of each transmitter position was de-multiplexed and the amplitude of a single element corresponding to the center of the transmitter was taken. Overall N measurements, each with N multiplexed samples, with a total of $N^2$ samples were performed and from the $N^2$ de-multiplexed signals only N signals, corresponding to N transmitter center were taken. This way, we recorded the response of each element in the virtual detector array to an identical input signal, mapping the sensitivity of the created virtual detection array. The measured sensitivity was used to normalize the results for the subsequent measurements. During the measurement, the transmitter and the mask were shifted seperatly in the x direction by 2 mm steps for the *N*=31 grid and by 1 mm steps for the *N*=59 grid.

### B. Acoustic field maps

The second set of measurements measures the acoustic field sampled by the mask detector and compares it to a single element detection. 2D radiation map of the transmitter was recorded by scanning the transmitter in the *y* direction with a 0.5 mm step size through a 150 mm range and for each transmitter distance, the mask was scanned in the x direction with 2 mm steps for the 31 elements mask the and 1 mm steps for the 59 elements mask. De-multiplexing the measurement data produced the 2D radiation map of the transmitter and the compatible signals in the time domain. To expand the span of the radiation map to a more visually comfortable grid, the measurement was repeated for two different *x* positions of the transmitter with the 59 elements mask and for four different *x* positions of the transmitter with the 31 elements mask, producing a complimentary set of radiation maps. Those maps were interlaced to produce a map with spatial sampling step of 0.5 in both, the x and the y directions for both masks. To measure the reference 2D radiation maps, two single-aperture masks were positioned in front of the center of the receiver. One with a single 1.5 mm aperture and the other with 1 mm aperture, thus emulating a point detector [20]. For each single aperture mask the transmitter was scanned in the *x*-*y* plane relatively to the detector with 0.5 step size in each direction. This reference measurement corresponds to the $\mathbf{W} = \mathbf{I}$ weighing matrix. The result of this measurement was a 2D radiation map of the transmitter recorded through raster scanning a single element detector and the compatible signals in the time domain.

### C. Detection isotropy

In the third set of measurement, the angular response of the virtual detector array was tested by manually rotating the transmitter with respect to the axis of the receiver, as illustrated in Fig. 3b. The measurement was performed in the far field with 220 mm distance between the transmitted and the detector. The CHA mask used for this measurement had apertures of 1 mm in diameter and the angle between the emitter and receiver (Fig. 3b) was rotated at several angles between 0º and 40º. Per angle a multiplexed measurement was performed and from the de-multiplexed signals only the central signal, pointed to the center of the transmitter was taken.

### IV. RESULTS

### A. Uniformity and sensitivity

Figures 4a and 4b respectively show the relative sensitivity of each element in the virtual detector arrays for the case of *N*=31 and *N*=59. The *x* axis in both figures, which represent the element index, was scaled by length to allow for a comparison

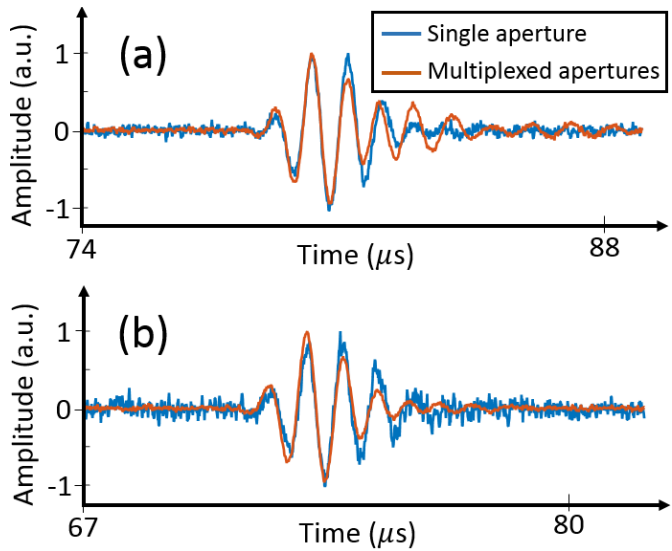

Fig. 5. Normalized waveform shape comparison of the strongest signal for a single vs. multiplexed aperture detection. The plots are for (a) an aperture with 1.5 mm diameter and 31 elements and (b) an aperture with 1mm diameter and 59 elements.

<div align="center">

TABLE II
SNR ADVANTAGE FOR MULTIPLEXED MEASUREMENTS

</div>

| Number of elements | Measured SNR Advantage | Calculated SNR Advantage |
|---|---|---|
| 31 | 2.94 | 2.87 |
| 59 | 3.99 | 3.91 |

*SNR gain achieved by the multiplexed measurement in comparison to the theoretical values.*

between the results. The span of the receiver with respect to the array elements is shown in the figure. For both masks, the length of the receiver, which was 38 mm, was smaller than lengths of the virtual detector array, which were 61.5 mm (Fig. 4a) and 59 mm (Fig. 4b). As both figures clearly show, the response of the virtual detectors dropped for indices outside the receiver span. While one might expect that all the virtual detectors outside the span of the receiver would receive a zero signal, the results in Figs. 4a and 4b depict a gradual decline in sensitivity outside the receiver span. This result may be attributed to diffraction: Because the aperture diameter is comparable to the acoustic wavelength, the transmission through the aperture is semi-isotropic [20]. It is therefore expected that some of the acoustic radiation from array elements not covered by the receiver would still reach the receiver.

### B. Acoustic field maps

Figures 5a and 5b respectively show the acoustic waveforms for the case of $N=31$ and $N=59$ in comparison to the waveforms obtained in the direct measurement with the compatible single aperture masks. In both cases, a good agreement was achieved between the signals of the multiplexed and direct measurement, where the noise level in the multiplexed measurement was lower. The expected and measured multiplexing SNR advantages for both masks are summarized in Table II, where the minute differences between the two may be attributed to production variations in the aperture sizes between the CHA and single-aperture masks.

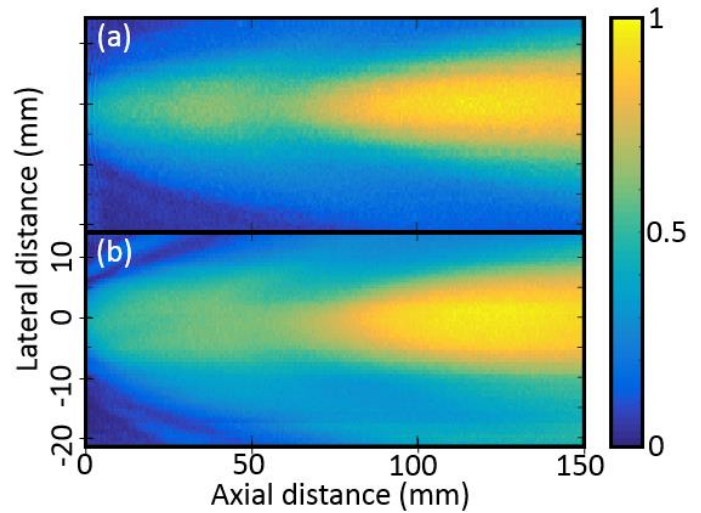

Fig. 6. Diffraction pattern of a round single element transmitter. Measured by a virtual detector with 1.5 mm diameter. Acquisition performed by (a) a single scanned detector and (b) multiplexed virtual array with 31 elements. The color bar represents the magnitude of the acoustic fields in arbitrary units.

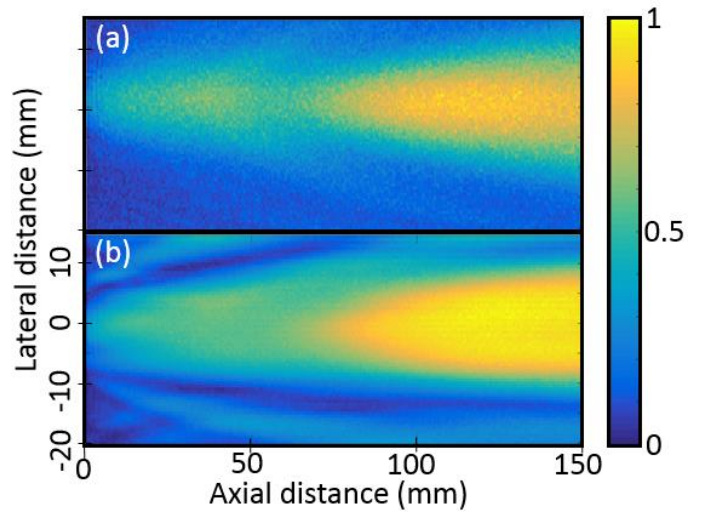

Fig. 7. Diffraction pattern of a round single element transmitter. Measured by a virtual detector with 1 mm diameter. Acquisition performed by (a) a single scanned detector and (b) multiplexed virtual array with 59 elements. The color bar represents the magnitude of the acoustic fields in arbitrary units.

Figures 6a and 6b show the 2D radiation map obtained with the CHA mask for the case of $N=31$ and with the corresponding single-aperture masks. As expected, the same radiation pattern was measured using both techniques, where the SNR in the CHA measurement was higher. Figure 7a shows the same radiation map characterized with a mask with $N=59$, where Fig. 7b shows the result obtained with the corresponding single-

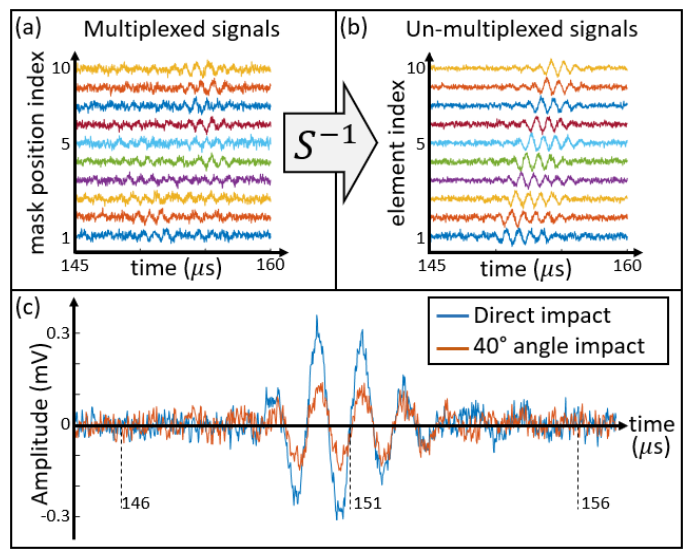

Fig. 8. A signal, hitting the detector from 40° angle. (a) shows the original signals, acquired by the multiplexed CHA array and (b) shows the un-multiplexed signals, both in (a.u.). (c) shows the amplitude comparison between a single signal detected in a direct impact and a single signal from a 40° angle impact signals.

### TABLE III
### ANGULAR SENSITIVITY

| Impact Angle [º] | Loss for unmasked detector [dB] | Loss for masked detector [dB] |
|---|---|---|
| 10 | 34.2 | 0.1 |
| 30 | 41.1 | 4.9 |
| 40 | 45.4 | 6.5 |

*Loss in comparison to a direct impact signal per impact angle for the signals acquired by the full-size detector and by a detector masked with the N=59 grid.*

aperture mask. The small differences between the radiation maps of Fig. 6 and 7 may be explained by the different aperture diameters used: 1.5 mm in Fig. 6 and 1 mm in Fig. 7. In terms of SNR, it may be visually appreciated that the SNR gain due to the multiplexed measurement is larger in Fig. 7 than in Fig. 6, in agreement with the quantitative results of Table I.

#### C. Detection isotropy

Fig. 8 summarizes the results of the acoustic measurement performed with $\theta =40°$ angle (Fig. 3b) using the CHA mask with $N=59$. Figure 8a shows the raw signals obtained for the first ten positions of the mask in the scan, whereas Fig. 8b shows the signals obtained after de-multiplexing for ten neighboring positions in the virtual detector array. As a result of the oblique incidence angle, the delay in the de-multiplexed signals increased with the position of the virtual detector in the array (Fig. 8b). In contrast, in the raw measurement (Fig. 8a), one cannot identify a clear structure of signal delays since each of the raw waveform represents a sum of the signals shown in Fig. 8b.

Fig. 8c compares the de-multiplexed signal of Fig. 8b to the corresponding de-multiplexed signal obtained with the same mask, but with a normal incidence angle, i.e. $\theta = 0$ (Fig. 3a). The figure shows a good correspondence between the two signals, where the main difference is that for the oblique incidence angle the signal was attenuated by 6.5 dB. In Table III, we summarize the attenuation values obtained for the CHA with respect to the normal incidence level and compare them to the values obtained when no mask was used, which represented the angular sensitivity of the receiver alone. As the Table shows, the large area of the receiver led to a very narrow angle of acceptance with a very high attenuation even for $\theta = 10°$. In contrast, when the CHA mask was used, broad angular acceptance was achieved.

acknowledgment

### V. CONCLUSION AND DISCUSSION

In conclusion, we developed and experimentally demonstrated a novel approach for mapping ultrasound fields using a CHA mask and a single element large-area detector. The CHA mask is composed of $N$ cells, where each cell may either transmit or block the acoustic wave that impinges on it. The signal detected by the receiver is thus a sum of $N$ acoustic signals, each weighted by either "1" or "0". By linearly translating the mask, a set of $N$ multiplexed signals is produced. The raw measurement data is then un-multiplexed to recover the individual $N$ acoustic signals at the location of each of the mask cells. Thus, the proposed scheme enables one to transform a single receiver into a virtual detector array.

To achieve the optimal SNR in our scheme, the binary codes produced on the CHA mask were derived from a cyclic S-matrix. The experiments were performed with two different masks, each with a different size, spacing and element count of $N=31$ and $N=59$, demonstrating the geometrical flexibility this technique offers. For both CHA masks, the de-multiplexed signals were in agreement with the signals obtained via a direct measurement by a single-aperture mask and with the theoretical multiplexing SNR gain of $\sqrt{N}/2$. In terms of angular sensitivity, our measurements show that CHA masks can achieve semi-isotropic detection similar to the one previously demonstrated for a single-aperture mask array [20].

We note that the results presented in this work are applicable for transducers that are used only in receive mode. While in conventional pulse-echo imaging the transducers are used as transceiver, where the same transducer is used to both generate and detect ultrasound, in other applications there is a separation between the generation and detection of ultrasound. For example, in UCT different transducers are used for the generation and detection of ultrasound, whereas in OAT ultrasound generation is performed optically. Future applications of CHA masks to OAT will require making several modifications to the proof-of-concept demonstrations described in this paper. First, while in our experiments most the detector area was blocked, thus limiting sensitivity, using CHA masks in an OAT system would require using the entire detector area to maximize the SNR. Second, since OAT is often performed at frequencies above 1 MHz, CHA masks for higher frequencies will be needed. In such masks both the thickness and aperture size will need to be scaled with the acoustic wavelength. Additionally, using CHA at higher frequencies, and accordingly shorter wavelengths, would require positioning the mask closer to the mask to minimize diffraction in the propagation from the mask to the detector. Alternatively, if the required distances become too small for the alignment tolerance of the mask, modeling of the signal distortion due to diffraction would be required to be included in the inversion procedure. Finally, we note that while the ultrasound signals detected in our experiments had a limited bandwidth, optoacoustic signals are naturally wideband. Nonetheless, OAT is often performed with piezoelectric detectors with a limited bandwidth, effectively reducing the measurement to a single acoustic band. Thus, the use of a limited-bandwidth transmitter in our experiments corresponds to merely a signal apodization within the effective band an OAT measurement.

One of the advantages of using CHA masks for ultrasound detection is that it adds new possibilities to OAT system design. First, one may use CHA masks to increase the number detectors in 3D OAT by transforming each of the array elements into its own virtual array. While the complexity of array fabrication and data acquisition generally limits 3D OAT systems to 512 elements [29],[30], using CHA masks could potentially increase the number of detectors beyond $10^4$, leading to more detailed images. While the same number of effective detectors may be achieved by merely scanning the array, it would not lead to the SNR advantage or increase acceptation angle offered by the CHA mask. Second, while producing arrays with a pitch below 100 μm is technologically challenging, limiting the lateral resolution of OAT, producing masks with apertures smaller than 100 μm represents a much small technical challenge. Thus, CHA masks may enable a new generation of high-resolution OAT systems.

## VI. Acknowledgment

The research reported in this article was partially supported by the Ollendorff Minerva Center.